\documentclass[12pt]{iopart}

\usepackage{graphicx}
\usepackage{dcolumn}
\usepackage{bm}
\usepackage{bbm}
\usepackage{epsfig}
\usepackage{mathrsfs}
\usepackage{stmaryrd}
\usepackage{color}
\usepackage{dsfont}
\usepackage{iopams}
\usepackage{subeqn}
\usepackage{hyperref}
\usepackage{amsfonts, amssymb}
\usepackage{pifont} 
\usepackage{subfigure}

\begin{document}

\title{Cavity-induced phase stability to decelerate a fast molecular beam via feedback-controlled time-varying  optical pumps}
\author{Zhihao Lan}
\address{SUPA, Department of Physics, Heriot-Watt University, Edinburgh EH14 4AS, UK}
\ead{lanzhihao7@gmail.com}
\author{Weiping Lu}
\address{SUPA, Department of Physics, Heriot-Watt University, Edinburgh EH14 4AS, UK}

\date{\today}

\begin{abstract}
We have identified a novel phase stability mechanism from the intracavity field-induced self-organization of a fast-moving molecular beam into travelling molecular packets in the bad cavity regime, which is then used to decelerate the molecular packets by feedback-controlled time-varying laser pumps to the cavity. We first applied the linear stability analysis to derive an expression for this self-organization in the adiabatic limit and show that the self-organization of the beam leads to the formation of travelling molecular packets, which in turn function as a dynamic Bragg grating, thus modulating periodically the intracavity field by superradiant scattering of the pump photons. The modulation encodes the position information of the molecular packets into the output of the intracavity field instantaneously. We then applied time-varying laser pumps that are automatically switched by the output of the intracavity field to slow down the molecular packets via a feedback mechanism and found that most of the molecules in the molecular packets are decelerated to zero central velocity after tens of stages. Our cavity-based deceleration proposal works well in the bad cavity regime, which is very different from the conventional cavity-based cooling strategies where a good cavity is preferred. Practical issues in realizing the proposal are also discussed.\end{abstract}

\pacs{37.10.Mn, 37.10.Vz, 37.30.+i }

\maketitle
\section{Introduction.}
\label{s1}

Recent progress in developing methods to produce, trap and control ultracold atomic gases has led to remarkable achievements, including the generation of atomic Bose-Einstein-condensates (BECs) \cite{r1}, the observation of degenerate Fermi gas  \cite{r2} with subsequent investigation of BEC-BCS (Bardeen-Cooper-Schrieffer) crossover  \cite{r3} and the realization of superfluid-to-Mott-insulator transition  \cite{r4}. Prompted by these successes, researchers are now attempting to enlarge the list of cold matters to include gaseous molecules  \cite{r5}. The ability to cool molecules below millikelvin temperatures promises to have a great impact on the fields of both physics and chemistry. Ultracold molecules offer the possibility to study exotic quantum phases through long-range and anisotropic electric dipole-dipole interactions  \cite{r6}. Trapped ultracold molecules could also be a suitable candidate for qubits in quantum computation  \cite{r7}, used to constrain the time variation in fine structure constant  \cite{r8}, search for parity violation  \cite{r9} and to test physics beyond the standard model  \cite{r10}. Moreover, the capability to create ultracold molecules and subsequently trap them in external electric and magnetic fields allows long interaction and interrogation times and therefore high-resolution spectroscopic measurements. Such measurements have already been carried  \cite{r11, r12} on radiative lifetimes of OH and NH in the vibrationally excited states. Ultracold molecules are also anticipated to be important in chemistry, as resonance and tunnelling phenomena could be dominating effects at ultracold temperatures, with reaction rates predicted to be many orders of magnitude larger than at room temperature for some species  \cite{r13}.

The conventional method to cool atoms uses many consecutive absorption emission cycles in a closed multilevel system to extract kinetic energy from the atoms. However, the method cannot generally be applied to molecular species, except for a few special cases \cite{r14}, because of their complex energy structure that precludes closed-cycling transitions. Ultracold molecules can be created from association of laser-cooled atomic species by photoassociation or on magnetic Feshbach resonances at microkelvin temperatures \cite{r15, r16}. Progress has been made recently to transfer these molecules in high vibrational levels to absolute ground states \cite{r17}.
These methods are nevertheless limited to atoms that can be laser cooled. Buffer gas cooling \cite{r18} is a general method, which can dissipatively cool complex molecular species by the use of thermalizing collisions with buffer gas in a cryogenic cell. The first buffer gas-cooled BEC has been reported recently \cite{r19}. Optical cavity cooling is another general scheme independent of the specific internal energy structure of the particles. It cools particles by a dissipative optical dipole force arising from the nonadiabatic dynamics between the optical field and particles in the cavity \cite{r20, r21, r22, r23, r24, r25, r26, r27, r28}. An advantage of this method is the low temperatures it can achieve, which are limited by the cavity linewidth and can be much lower than the Doppler limit set by the atomic linewidth. Optical cavity cooling of atoms has been demonstrated experimentally by several research groups \cite{r29, r30, r31}. In general, cavity cooling is a secondary cooling scheme that is employed to reduce further the temperature of a primary cold sample, typically of the order of tens to hundreds of millikelvins, to the ultracold regime at submillikelvin temperatures.

The primary cold molecular sample can be obtained via the phase space filtering technique, where conservative electrostatic \cite{r32, r33}, magnetic \cite{r34, r35} or optical potentials \cite{r36} are used to filter out a narrow velocity distribution of a hotter molecular gas and then transfer it from the moving frame in the molecular beam to zero velocity in the laboratory frame. Different from the usual cooling methods, which focus on the condensation of stationary molecular ensembles in phase space \cite{r37}, the phase-space filtering technique focuses on the deceleration of a fraction of molecules in a fast molecular beam to near zero velocities. Electrostatic Stark deceleration is a well-developed scheme of this type that uses rapidly switched electrical fields to create a moving potential that traps and slows a subset of the initial molecular distribution. A gas of $10^6$cm$^{-3}$ polar molecules in a single quantum state at 10 mK was reported \cite{r32}. Following the success of the electrostatic Stark deceleration, other deceleration schemes have recently been studied theoretically. A microwave Stark decelerator was proposed to slow a hot polar molecular beam by using a time-varying standing wave in a cavity that is created by timing the external pump source in a similar way to the electrostatic Stark decelerator \cite{r38}. Deceleration of a particle in a bistable optical cavity is another scheme in which the deceleration force is induced by feedback-controlled switching of the optical pumps between a high and a low state \cite{r39}. A setback of this scheme is that the cooling effect seems to be washed out quickly with increasing particle number due to lack of collective motion of particles in the cavity. Most recently \cite{r40}, we have shown that a spatially homogeneous supersonic molecular beam travelling along the axis of a low-finesse optical cavity can undergo a phase transition to spatially periodic travelling molecular packets with strong collective motion. The travelling molecular packets, in turn, function as a dynamic Bragg grating that scatters the pump fields superradiantly to form an intracavity field. The intracavity field then switches on and off dynamically as the travelling molecular packets move along each cycle of the cavity mode. The nonadiabatic nature of the cavity setup gives rise to a friction force, which slows most of the molecules in the beam to zero central velocity.

Whereas the decelerator in \cite{r40} operates in the intermediate cavity regime where the two ingredients of an effective deceleration process, phase stability and deceleration force, need to be balanced with each other; in this paper, we present a new scheme for decelerating a fast molecular beam in the bad cavity regime where phase stability and deceleration force can be engineered separately. Specifically, firstly, unlike the traditional phase stability mechanism in electrostatic Stark deceleration, where the stability realized has been imposed by an external source, the phase stability mechanism in our scheme emerges spontaneously from the cavity-induced collective behaviour of all the molecules in the beam and works well in the bad cavity regime. Secondly, unlike the deceleration force in previous studies that comes from the nonadiabatic dynamics in the good cavity regime, the deceleration force in this paper arises from the automatic switching of the pump intensity between two levels via a feedback loop controlled by the output of the intracavity field, where the effective operation of the feedback loop also relies on a bad cavity, thus allowing us to engineer them separately.

We note that while the deceleration studies usually focus on the slowdown of a fast molecular beam, where the velocity spread is heated as in \cite{r40} due to the nonadiabatic widening at the end stage, the heating effect is largely suppressed in the present scheme (see figures \ref{fig6} and \ref{fig7}) compared with \cite{r40} due to the adiabatic nature of the cavity setup, allowing us to maintain a low transverse temperature and high density of the beam. Moreover, conventional cavity cooling studies \cite{r20, r21, r22, r23, r24, r25, r26, r27, r28} focus on the nonadiabatic physics in the good cavity regime, whereas our work uncovers the new physics hidden in the bad cavity regime and successfully extends the feasibility of cavity-assisted obtaining of ultracold or cold molecular samples from the good to intermediate to bad cavity regime.

The paper is organized as follows. In section \ref{s2}, we present the optical cavity-based deceleration scheme, in which we introduce two different descriptions, discrete and statistical, for the convenience of modelling and analysis. In section \ref{s3}, we first apply the linear stability analysis to derive an expression for the phase transition of a spatially homogeneous fast molecular beam in the adiabatic limit. We further show in detail that the phase transition of the beam leads to the formation of a self-consistent molecule-field steady state, which forms the base of the proposed deceleration scheme in the subsequent sections. The time-varying feature of the pump intensity for decelerating the molecular packets is presented in the first part of section \ref{s4}, which is then followed by numerical simulations. The composite deceleration scheme in section \ref{s5} shows that the deceleration proposal in the present study is complementary to the one in \cite{r40}. However, if the readers bypass this section, this will not prevent their grasp of the main idea of the present study. In section \ref{s6}, we discuss in detail the practical issues in realizing the proposed deceleration method. We present our main conclusions in section \ref{s7}.

\section{The model}
\label{s2}

We consider a fast molecular beam entering (nearly axial) an optical cavity that supports a standing-wave mode of the form $\cos(kx)\exp(-i\omega_c t)$, where $k$ is the wavenumber and $\omega_c$ the bare cavity resonance frequency, see figure \ref{fig1}. The cavity is pumped transversely by laser beams far-off-resonance from any electronic transitions of the molecules with pump frequency $\omega_p$. At low saturation where the spontaneous emissions of the molecules are negligible, we can adiabatically eliminate the internal dynamics of the molecules and treat them as classical polarizable point objects. Therefore, the deceleration method studied in this paper is in principle applicable to a wide class of species, ranging from atoms to molecules or even to nanoparticles. A moving molecule inside the cavity serves as the effective refractive index that shifts the cavity resonance frequency in a position-dependent way, $U(x)=U_0\cos^2(kx)$, where $U_0=-\omega_p\rm{Re}\left[\alpha(\omega_p)\right]/(\epsilon_0V)$ \cite{r26, r39}, $\alpha(\omega_p)$ is the polarizability of the molecule, $V$ the mode volume of the cavity and $\epsilon_0$ the permittivity of free space. The scattering loss resulting from the imaginary part of $\alpha(\omega_p)$ has been neglected due to the far-off-resonance scheme. In the semiclassical limit, the combined system dynamics of the intracavity field amplitude and the centre-of-mass motions of the molecules can be described by the following coupled differential equations in one dimension \cite{r26}:
\begin{figure}
\begin{center}
\resizebox{1.0\columnwidth}{!}{\includegraphics{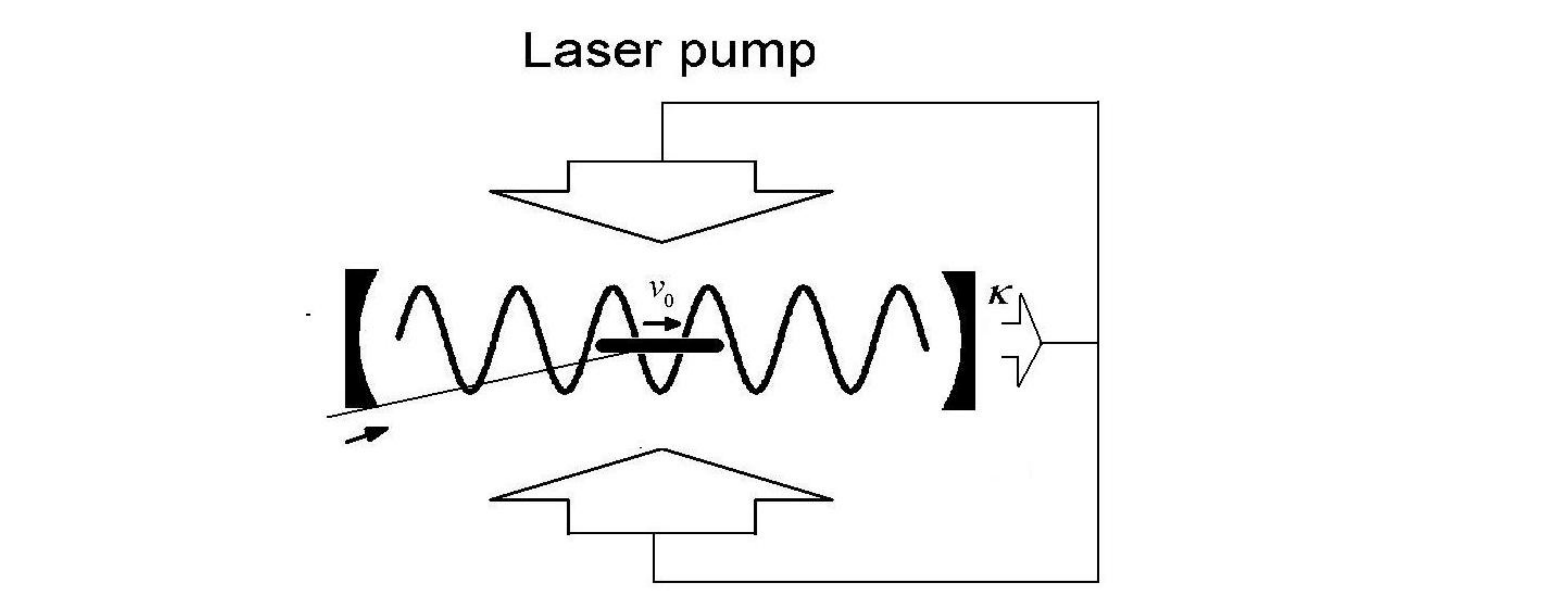}}
\end{center}
\caption{ Schematic of an optical cavity based molecular decelerator with feedback-controlled time-varying laser pumps.} 
\label{fig1} 
\end{figure}
\begin{eqnarray}
\dot{\alpha}=(i\Delta_c-\kappa)\alpha-iU_0\sum_j\cos^2(kx_j)\alpha+i\eta\sum_j\cos(kx_j), \nonumber \\
\ddot{x}_j=\frac{\hbar k}{m}\left[ U_0|\alpha|^2 \sin(2kx_j)-2\eta \textrm{Re} (\alpha )\sin(k x_j) \right],
\label{eq1}
\end{eqnarray}
where $\alpha$ is the amplitude of the photon number in the cavity, $\Delta_c=\omega_p-\omega_c$ the detuning of the pump lasers with respect to the cavity resonance, $\kappa$ the cavity decay rate, $x_j$ and $m$ the position and mass of the $j$th molecule, $j=1, 2, \cdots, N$, where $N$ is the total number of molecules and $\eta$ the effective pump amplitude of a molecule. Note that this study adopts the far-off-resonance description by using the polarizability of the molecule $\alpha(\omega_p)$, which in principle applies to all polarizable particles, instead of the near-resonance description used in \cite{r40} and the noise terms as in \cite{r40} are also neglected, since we found in the present case where the molecular beam is fast-moving, the noise terms have a negligible effect on the system dynamics, which is supported by relevant work as in \cite{r26}.

When the molecular number in the beam is sufficiently large, the molecules can
be	described	statistically	by	the	position	and	velocity	distribution	function	$f (x, v, t)$
\cite{r28},	for	which	the	position-related	summations	can	be	reexpressed	as
$\sum_j\cos(kx_j)\rightarrow N\int f(x,v,t)\cos(kx)dxdv$ and $\sum_j\cos^2(kx_j)\rightarrow N\int f(x,v,t)\cos^2(kx)dxdv$. 
Consequently, the first equation in equation (\ref{eq1}) can be rewritten as 
\begin{equation}
\hspace{-2.5cm} \dot{\alpha}=(i\Delta_c-\kappa)\alpha-iNU_0\alpha\int f(x,v,t)\cos^2(kx)dxdv+iN\eta\int f(x,v,t)\cos(kx)dxdv 
\label{eq2}
\end{equation}
and the distribution function $f(x,v,t)$ obeys the collisionless Boltzmann equation
\begin{equation}
\frac{\partial f(x,v,t)}{\partial t}+v\frac{\partial f(x,v,t)}{\partial x}+\frac{F(x,t)}{m}\frac{\partial f(x,v,t)}{\partial v}=0
\label{eq3}
\end{equation}
where $F(x, t)$ is the force exerted on the molecules:
\begin{equation}
F(x,t)=\hbar kU_0|\alpha|^2\sin(2kx)-2\hbar k \eta \textrm{Re}(\alpha)\sin(kx)
\label{eq4}
\end{equation}
Equations (\ref{eq2})-(\ref{eq4}) form the statistical description of the system. The above two descriptions
serve different purposes in this paper. While the statistical description is mainly used for theoretical analysis, the discrete description is used for numerical simulations directly. We note that the statistical description is based on the method of collisionless Boltzmann equation we developed recently \cite{r28}. Similar methods for phase-space distribution calculations of classical \cite{r41} and quantum \cite{r42} gases were developed also by other researchers recently. This work is closely related to the Vlasov approach, the technical details of which are given in \cite{r41} and references therein.

\section{Cavity-induced self-organization of the fast molecular beam}
\label{s3}

When a stationary cold atomic cloud is placed in a standing-wave cavity pumped by a laser in a direction perpendicular to the cavity axis, a phase transition has been predicted from the initial homogeneous atomic distribution to a regular patterned state that maximally scatters the pump photons into the cavity by atomic crystallization at either the even or the odd antinodes of the cavity mode \cite{r23, r24}. The phase transition occurs above a certain pump intensity (threshold) and spontaneously breaks a discrete translational symmetry of the system. When a fast-moving molecular beam is considered instead of the stationary cold atomic cloud, a new parameter, the central velocity of the beam $v_0$, is introduced into the system, which brings in new physics that has no counterpart as in the case of the stationary cold atomic cloud. The phase transition of a fast-moving gas beam in a ring cavity pumped by two counter-propagating laser fields through the cavity mirrors has been studied \cite{r43}. It is shown to occur only when the frequency shift induced by the particles is larger than the cavity linewidth, which implies a large ensemble of particles or a high-$Q$ cavity. In the ring cavity, which is the simplest multimode cavity supporting two counter-propagating modes, the locations of the antinodes are collectively determined by the particles moving in the cavity, instead of self-emergent as in the standing-wave cavity, so the translational symmetry breaking of the system is continuous rather than discrete as in the standing-wave cavity. This collective determination of the antinodes results in a shift of the peak density position of the particles from the optical field minima in the cavity and thus the system cannot reach a time-independent self-consistent particle-field steady state \cite{r43}. In the present study, we expect a self-consistent molecule-field steady state because of the use of a standing-wave cavity where the antinodes are fixed by the cavity geometry and such a steady state is also expected to be achieved in the bad cavity regime where there is no dissipative factor to destroy its stability. This steady-state solution is the base for a new phase stability mechanism suitable for multistage deceleration in our system.

In this section, we will study the dynamics of a fast molecular beam in the standing- wave cavity and derive analytically in the adiabatic limit the threshold for a phase transition and the scaling laws of the dynamics with respect to the parameters of the system. Numerical investigations will then be carried out to study the dynamical interplay between the moving molecules and the intracavity field, particularly focusing on the self-consistent moleculeÐfield steady state.

\subsection{ Linear stability analysis and phase transition}
\label{s31}

We consider the phase transition as a linear stability problem of the solution of the coupled intracavity field and Boltzmann equations (\ref{eq2})-(\ref{eq4}). In obtaining the threshold pump for the onset of the phase transition in the adiabatic limit, we linearize the coupled equations around the trivial solution (initial conditions) and then solve the linearized equations as an eigenvalue problem. The parameter dependence on the threshold gives the scaling laws of the system. To do so, we expand the variables
\begin{equation}
\alpha(t)=\alpha_0+\delta\alpha(t), \hspace{1cm} f(x,v,t)=f_0+\delta f(x,v,t),
\label{eq5}
\end{equation}
where $\alpha_0$ is the initial photon number amplitude in the cavity, and $f_0$ is the initial position-velocity distribution function of the molecular beam, assumed to be uniform in space and Gaussian in velocity, $f_0=f_x \cdot f_v\equiv1/L\cdot \textrm{exp}[-(v-v_0)^2/2\sigma^2)]/\sqrt{2\pi\sigma^2}$, where $v_0$ is the central velocity, $\sigma$ the velocity spread and $L$ the length of the beam, respectively. By substituting (\ref{eq5}) into equations (\ref{eq3}) and (\ref{eq4}) and keeping only linear terms, we obtain the linearized Boltzmann equation
\begin{equation}
\frac{\partial \delta f}{\partial t}+v\frac{\partial \delta f}{\partial x}-\frac{2\hbar k \eta}{m} \textrm{Re}(\delta\alpha)\sin(kx)\frac{\partial f_0}{\partial v}=0
\label{eq6}
\end{equation}
We then express the trial solution of equation (\ref{eq6}) in the form of a travelling density wave with velocity $v_0$, i.e.

\begin{equation}
\delta f=e^{\lambda t} f_v \left[ A \sin(kx-kv_0t)+B\cos(kx-kv_0t)\right]
\label{eq7}
\end{equation}
where $A$ and $B$ are constants and $\lambda$ is to be determined by the system parameters. This trial solution of the travelling first harmonic wave with velocity $v_0$ is based on two facts: the beam is travelling with central velocity $v_0$ and the source term in its parent equation (\ref{eq6}) is a first harmonic wave. For convenience, we recast the trial solution as
\begin{equation}
\delta f=e^{\lambda t}\left[ A(t)\sin(kx)+B(t)\cos(kx)\right]
\label{eq8}
\end{equation}
where $A(t)=A\cos kv_0t+B\sin kv_0t$ and $B(t)=B\cos kv_0t-A\sin kv_0t$ are two orthonormal bases. In the adiabatic limit, the intracavity field follows the change in the molecular distribution instantaneously, and so $\textrm{Re}(\delta\alpha)=-e^{\lambda t} N \eta \delta_c B(t)/\left[ 2(\delta_c^2+\kappa^2)\right]$ from equation (\ref{eq2}), where $\delta_c=(\Delta_c-U_0N/2)$ is the modified cavity detuning. Substituting the trial solution (\ref{eq8}) together with the expression for $\textrm{Re}(\delta\alpha)$ into equation (\ref{eq6}), we obtain
\begin{eqnarray}
\lambda e^{\lambda t}\left[ A(t)\sin kx+B(t) \cos kx\right] +e^{\lambda t}kv_0\left[B(t)\sin kx -A(t)\cos kx \right] \nonumber \\
+e^{\lambda t}kv \left[ A(t)\cos kx-B(t)\sin kx \right] -e^{\lambda t} k(v-v_0)\rho [B(t)] \sin kx =0
\label{eq9}
\end{eqnarray}
where $\rho=\hbar N \delta_c \eta^2 /\left[m\sigma^2(\delta_c^2+\kappa^2)\right]$. The two Fourier components in equation (\ref{eq9}), $\sin kx$ and $\cos k x$, must equal zero separately, which leads to the eigenvalue equation
\begin{equation}
\left[
\begin{array}{cc} 
\lambda & kv_0-kv-\rho k(v-v_0) \\
kv-kv_0 &\lambda    \end{array} \right] \left[ \begin{array}{c}  A(t)  \\ B(t) 
\end{array} \right]  =0
\label{eq10}
\end{equation}

\noindent
with the solutions $\lambda^2=\left[k(v-v_0)\right]^2(-\rho-1)$. When $\lambda>0$, i.e.
\begin{equation}
\eta>\eta_{\rm{thr}}\equiv \sqrt{\frac{m}{\hbar}}\frac{\sigma}{\sqrt{N}}\sqrt{\frac{(\delta_c^2+\kappa^2)}{(-\delta_c)}}
\label{eq11}
\end{equation}
the trivial solution becomes unstable (phase transition), which leads to the exponential growth of a travelling density wave in the form of equation (\ref{eq7}) until saturation occurs from the nonlinear effects. Expression (\ref{eq11}) defines the threshold pump for the phase transition and also gives the scaling laws with respect to the parameters of the system. We note that in the adiabatic limit, where the intracavity field always follows the molecular motion instantaneously, the threshold
is independent of the central velocity of the beam but proportional to the velocity spread (the
adiabatic condition is discussed in detail in section \ref{s6} ). Also as $\eta_{\rm{thr}}^2\propto 1/N$ , the phase transition
is more likely to be observed for a large ensemble of molecules. Equation (\ref{eq11}) is consistent with the mean-field approximation \cite{r24, r26} and our previous work \cite{r28} under the relation $m\sigma^2 =k_BT/2$.

The physical mechanism underlying the self-organization of the fast molecular beam in the adiabatic limit is similar to that for a stationary cold atomic cloud in standing-wave cavity \cite{r23, r24}. The travelling molecules being transversally pumped by the lasers scatter photons into the cavity according to the source term $i\eta\sum_j\cos(kx_j)$ in the first equation of (\ref{eq1}). Molecules in the nodes of the standing-wave cavity mode do not make a contribution, whereas those in the antinodes scatter maximally. The photons scattered by molecules separated by half a wavelength have opposite phase and interfere destructively, thus preventing the buildup of the intracavity field for the uniform spatial distribution of the beam. However, due to density fluctuations of the molecules, a small intracavity field can emerge momentarily, which, for red-detuned laser pumps, creates an attractive optical potential to pull molecules to every other antinode of the cavity mode. When the pump intensity exceeds a certain level (threshold), this induced molecular redistribution within wavelength-spaced wells at every other antinode can strongly enhance the Bragg-type scattering of the pump photons into the cavity, which in turn further deepens the optical potential and traps more molecules in a runaway process. In the initial stage, the modulation of the molecular distribution function grows exponentially in the form given in equation (\ref{eq7}), evidenced by direct simulations of equation (\ref{eq1}) as shown in figure \ref{fig2}(a), where the position-velocity plots in the initial stage are given in moments (1)-(4). The intracavity intensity also grows exponentially in this period as shown in figure \ref{fig2}(b), where the corresponding moments are also marked.
 
\subsection{Formation of travelling molecular packets}
\label{s32}

\begin{figure}
\begin{center}
\resizebox{1.0\columnwidth}{!}{\includegraphics{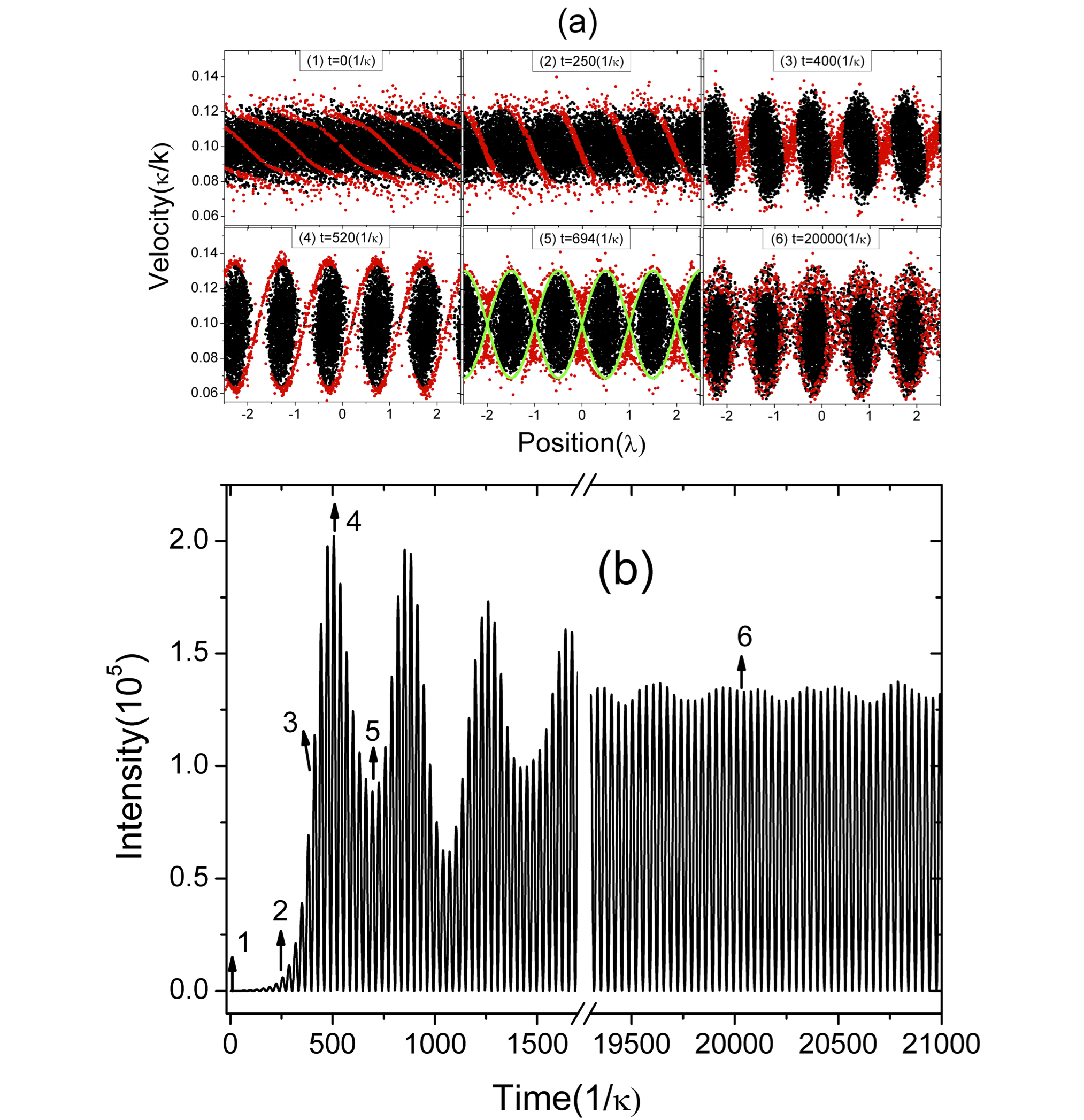}}
\end{center}
\caption{ (a) Phase-space plots of the molecular beam at different moments (1)-(6) from the simulation of equation (\ref{eq1}), where the corresponding intracavity intensities are marked in (b). The molecules marked with red are those outside the separatrix (green) at moment (5), which is determined by the intracavity intensity at this moment. These molecules are tracked to help understand the self-organization and evolution of the beam and the slow oscillation of the intensity profile. (b) The evolution of the intracavity intensity with time. Parameters used in the simulation $\hbar k^2/\kappa m=10^{-6}, U_0=-10^{-7}\kappa, \Delta_c=-10\kappa, N=10^4, \eta=2.5\eta_{\rm{th}}$  and the initial distribution of the beam is Gaussian in velocity, with $k v_0/\kappa=0.1, k\sigma/\kappa=0.01$ ,   and homogeneous in space within the length of five wavelengths (only three are shown (a)). Periodic boundary condition is used in the simulation.  } 
\label{fig2}
\end{figure}

The runaway process is eventually saturated by the nonlinear effects of the system when the amplitudes of the intracavity field and the travelling molecular wave have grown sufficiently strong (moment (4) in figure \ref{fig2}). Figure \ref{fig2} further shows the long-term time evolution of the molecular distribution and intracavity intensity, where the intensity exhibits two characteristic oscillations after the initial exponential growth (figure \ref{fig2}(b)). As we will see below, whereas the period of the fast oscillation corresponds to the time for the trapped majority of the molecules by a moving optical lattice to travel through a cycle of the standing-wave cavity mode, the slow oscillation is transient and related to the motions of the minor molecules that are untrapped by the moving lattice.

After saturation, the majority of the molecules are found to be bunched into packets
and move synchronically with central velocity $v_0$ in the cavity; thus in order to illustrate the
dynamics of the system, the distribution function of the molecules can be approximately written
as $f (x-v_0t)$. Under this approximation, the last term on the right-hand side of equation (\ref{eq2}) can
be rewritten as $N\int \cos(kx) f(x-v_0 t) dt = N \cos(kv_0t)\int \cos(kx)f(x)dx-N\sin(kv_0t)\int\sin(kx)f(x)dx 
=-N_{\rm{eff}}\cos(kv_0t)$ where we have set moment (4) as $t = 0$ such that the maximum values of the molecular position distribution $f (x)$ are positioned at $x=\cdots -3\pi, -\pi, \pi, 3\pi, \cdots$, so the integral $\int \sin(kx) f(x) dx=0$, and $N_{\rm{eff}} =-N \int \cos(kx)f(x)dx$ is the effective number of molecules. Since $NU_0\ll\kappa$ the second term in the right-hand side of equation (\ref{eq2}) and the first term in the right-hand side of equation (\ref{eq4}) can be neglected. Therefore, the important dynamics of the system in the adiabatic limit can be approximately expressed from equations (\ref{eq2}) and (\ref{eq4}) as
\begin{eqnarray}
|\alpha|^2=I_0\cos^2(kv_0t) \nonumber \\
m \ddot{x} =F_0\left[ \sin(kx+kv_0t)+\sin(kx-kv_0t)\right]
\label{eq12}
\end{eqnarray}
where $I_0=\eta^2 N_{\rm{eff}}^2/(\Delta_c^2+\kappa^2)$ is the amplitude of the intracavity intensity and $F_0=\hbar k\Delta_c N_{\rm{eff}}\eta^2/(\Delta_c^2+\kappa^2)$
 is the amplitude of the dipole force acting on the bunched molecular packets from the standing-wave potential, which consists of two counter-propagating optical lattices with the same velocity $v_0$ as the bunched molecular packets. As discussed in electrostatic Stark deceleration \cite{r33}, the lattice whose velocity comes close to the bunched molecular packets interacts more significantly with it, so the lattice propagating in an opposite manner to the bunched molecular packets can be neglected. As such, the important system dynamics of the bunched molecular packets moving in standing-wave cavity is reduced to bunched molecular packets travelling within an optical lattice of the same velocity. So the bunched dynamics of the molecular packets comes essentially from the trapped dynamics of the packets by the potentials of the moving lattice, equivalent to the transportation scheme as in Stark deceleration. The phase stability in our scheme thus results from the cavity-induced collective behaviour of all the molecules in the beam. This mechanism ensures the phase stability of the majority of the molecules in the cavity rather than a small fraction determined by the acceptance volume as in phase space filtering techniques.

After illustrating the trapped dynamics of the molecular packets by the moving lattice, we now turn to the intracavity intensity. The first equation of (\ref{eq12}) captures well the fast oscillatory behaviour of the intensity profile, which stems from the fact that bunched molecular packets travel along each cycle of the standing-wave cavity mode with period $\pi/kv_0$ and thus switch the intracavity intensity on and off dynamically with the same period $\pi/kv_0$. The slow oscillatory behaviour of the intensity profile is related to the motions of the minor molecules that are untrapped by the moving lattice. These untrapped molecules (marked with red in figure \ref{fig2}(a)) are best identified at the first dip of the intensity profile (moment (4) in figure \ref{fig2}(a)) when they move to the space between the bunched molecular packets. Molecules within the separatrix (green curve), which is determined from the height of the potential with the moving lattice, are trapped, while those outside the separatrix are untrapped. These untrapped molecules are then tracked at different moments (1)-(6) as shown in figure \ref{fig2}(a) to help understand the self-organization and evolution of the beam and the slow oscillatory behaviour of the intensity profile. Since the untrapped molecules are travelling within the moving lattice from one potential to another, when they travel to the crests (troughs) of the potential, which correspond to the minima (maxima) of the spatial molecular distribution, they mainly serve as ÔdefectsÕ (ÔgainsÕ) that scatter photons with opposite (same) phase to the trapped majority of the molecules and thus undermine (enhance) the intracavity intensity slightly. Since the dynamics of the intracavity intensity is determined mainly by the trapped molecular packets, the trajectories of the untrapped molecules are constantly modified by the intracavity field in an uncorrelated manner. Also the trapped and untrapped molecules near the interface of the separatrix can switch their roles as the intensity varies. As a result, the correlation between the untrapped molecules is eventually washed out, leading to the disappearance of the slow oscillatory behaviour in figure \ref{fig2}(b).

\begin{figure}
\begin{center}
\resizebox{1.0\columnwidth}{!}{\includegraphics{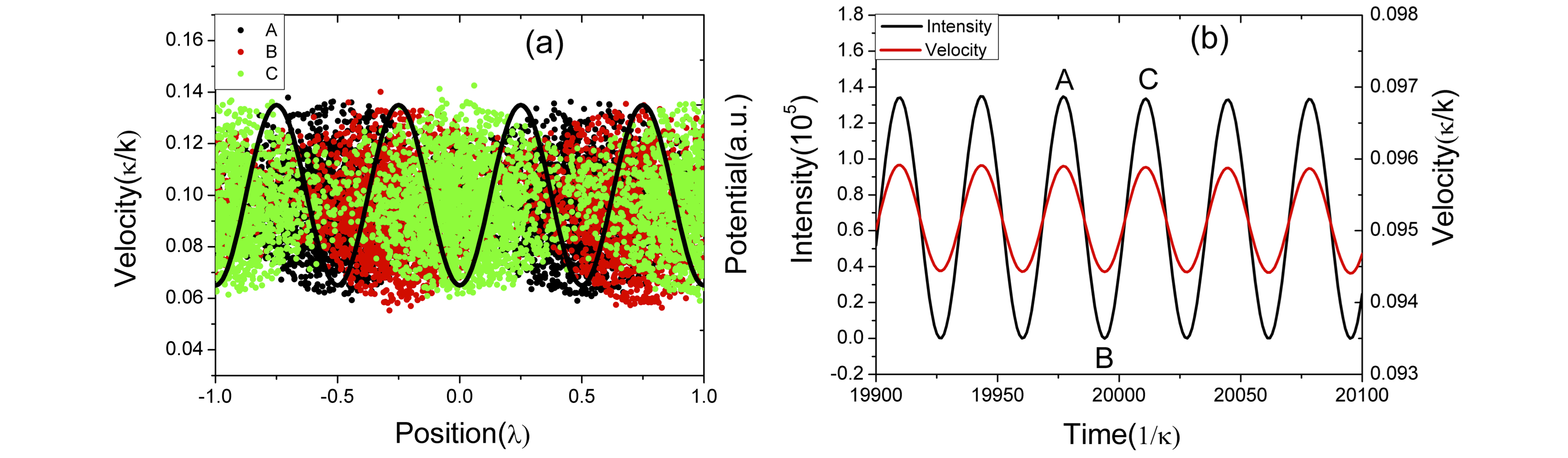}}
\end{center}
\caption{ Self-consistent molecule-field steady state of the system. (a) The phase-space plots of the molecular distribution at three moments A, B and C; also shown is the standing-wave potential proportional to $\cos^2(kx)$ . At moments A and C, the centres of the molecular packets are located at the troughs of the standing-wave potential, while at moment B, the centres are located at the crests. (b) The dynamical interplay between the intracavity intensity and the central velocity of the molecular packets.} 
\label{fig3}
\end{figure}

In the moving frame with the lattice, the trapped molecules are circulating approximately along closed orbits in phase space within the lattice potential, so the velocity distribution of the trapped molecules is determined by the amplitude of the lattice potential. An increase of the pump intensity will lead to an increase of the amplitude of the lattice potential, which in turn will widen the velocity distribution of the trapped molecules. In the same moving frame, the untrapped molecules are travelling along the lattice potential. The time required by the untrapped molecules to travel through a cavity mode is also determined by the amplitude of the lattice potential and decreases with increasing pump intensity, which accelerates the disappearance of the slow oscillatory behaviour of the intensity profile.

After the disappearance of the slow oscillatory behaviour of the intensity profile, the system approaches its self-consistent molecule-field steady state, the main characteristic of which is the fast oscillation of the intracavity intensity correlated with the travelling of bunched molecular packets along the standing-wave potentials ((6) in figure \ref{fig2}). At this stage, the system dynamics is more accurately described by equation (\ref{eq12}). In the above analysis, to illustrate the phase stability mechanism, we have neglected the relative minor effects of the optical lattice in the second equation of (\ref{eq12}) that has the opposite velocity to the bunched molecular packets. This lattice has, however, a discernible effect on the motion of bunched molecular packets, inducing a weak periodic oscillation to the central velocity of the travelling molecular packets (figure \ref{fig3}(b)). This weak oscillation is derived from the ascending and descending processes of the bunched molecular packets in the standing-wave potentials, as evidenced from figure \ref{fig3}, where from time A to B (or from time B to C), the molecular packets climb up (or down) the standing-wave potentials; thus the central velocity decreases (or increases). These main characteristics of the self-consistent molecule-field steady state form the foundation for the multistage deceleration to be discussed in the following sections.

\section{Deceleration scheme based on time-varying optical pumps}
\label{s4}
The self-consistent molecule-field steady state as discussed in the previous section implies a new phase stability mechanism, that arises spontaneously from the cavity-induced collective behaviour of all the molecules in the beam, unlike the phase stability mechanism of electrostatic Stark deceleration, which is imposed by external source. Since at the steady state the position information of the bunched molecular packets is encoded in the output of the intracavity intensity instantaneously, we can modulate automatically the pump intensity of the lasers by using the output of the intracavity intensity via the feedback mechanism, which will create a deceleration force to slow down the bunched molecular packets in a similar way to the electrostatic Stark decelerator. We first introduce the principle of our deceleration scheme in section \ref{s41}. Simulation results and analysis are presented in section \ref{s42}.

\subsection{The deceleration principle}
\label{s41}
Since the bunched molecular packets behave like a single molecule modulated by the effective number $N_{\rm{eff}}$, we use the single molecule model to illustrate the idea of deceleration. In the adiabatic limit, the intracavity field follows the motion of the molecule instantaneously, as given by equation (\ref{eq1}):

\begin{eqnarray}
|\alpha|^2=\frac{\cos^2(kx)\eta^2}{\kappa^2+\left[\Delta_c-U_0\cos^2(kx)\right]^2}=\frac{\eta^2}{K(x)} \nonumber \\
m\ddot{x}=\frac{\hbar k \Delta_c\eta^2\sin(2kx)}{\left[\Delta_c-U_0\cos^2(kx)\right]^2+\kappa^2}=-\frac{d}{dx}[V(x)],
\label{eq13}
\end{eqnarray}
where $K(x)=\left[ \kappa^2+(\Delta_c-U_0\cos^2(kx))^2\right]/\cos^2(kx)$ is the position-dependent intensity modulation parameter and $V(x)=-\hbar \Delta_c\eta^2\left[\rm{arctan}(\Delta_c/\kappa-U_0\cos^2(kx)/\kappa)\right]/U_0\kappa$ is the potential. As $\Delta_c \gg U_0$, we expand $\cos^2(kx)$ to its leading order and neglect the constant potential term, $V(x)\approx \hbar \Delta_c\eta^2\cos^2(kx)/(\kappa^2+\Delta_c^2)$. When the pump is constant (no feedback), the molecule travels along the cosine-squared conservative potential and there will be no net force to accelerate or decelerate the molecule.

\begin{figure}
\begin{center}
\resizebox{0.8\columnwidth}{!}{\includegraphics{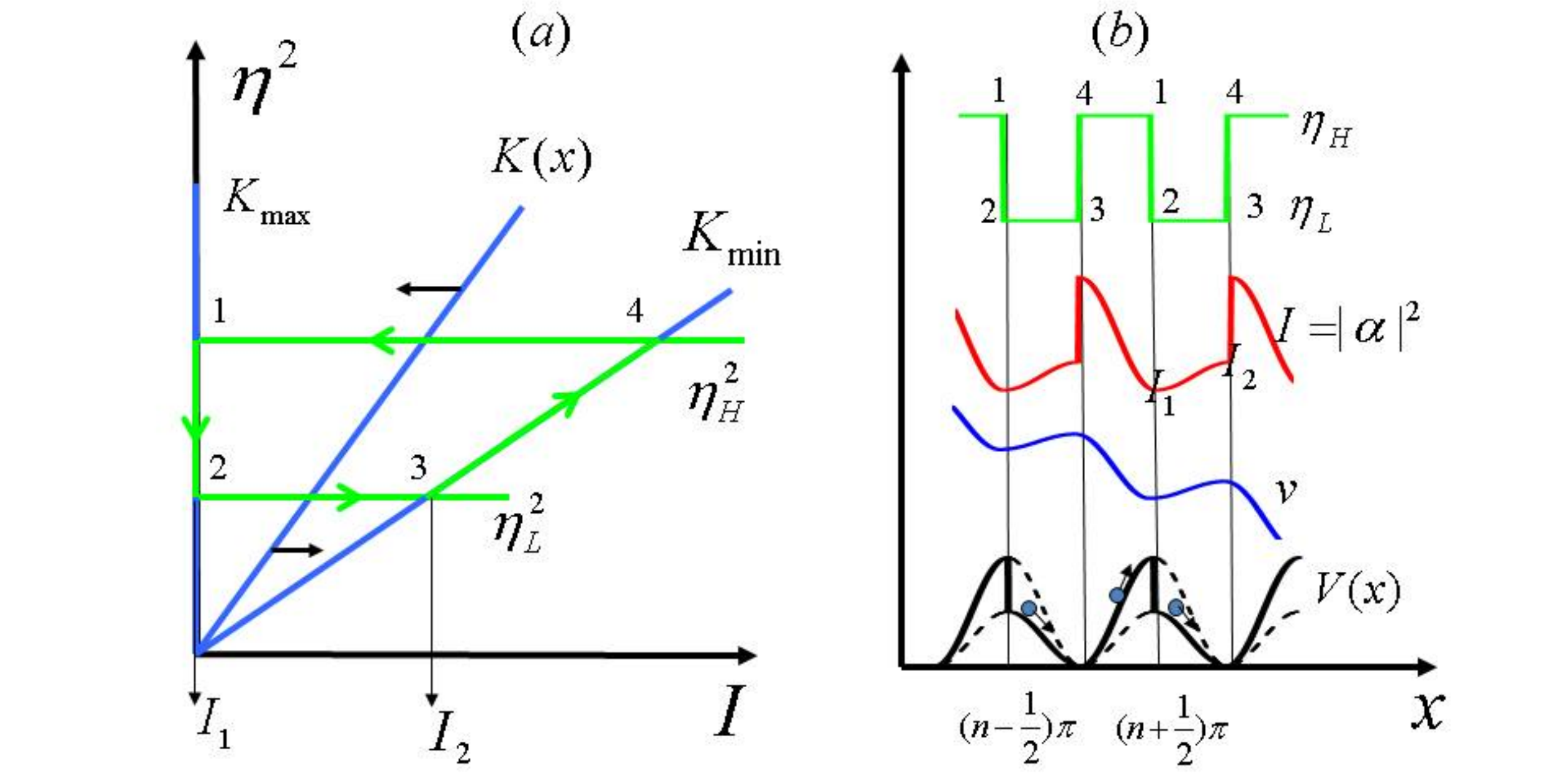}}
\end{center}
\caption{ The working principle of the optical cavity based molecular decelerator with feedback-controlled time-varying optical pumps. (a) The operation of the feedback loop. (b) The evolution of system quantities with the time-varying optical pumps.  } 
\label{fig4}
\end{figure}

Now we introduce the time-varying optical pumps in the following way. When the molecule is about to move down the potential hill as shown by point 1 in figure \ref{fig4}, the pump is switched from the high intensity level $\eta_{\rm{H}}$ to the low intensity level $\eta_{\rm{L}}$ (jump from point 1 to 2 in figure \ref{fig4}). The molecule gains kinetic energy during the moving-down process, which corresponds to the potential difference between points 2 and 3 in $V(x)$. The pump is then switched back to the high level $\eta_{\rm{H}}$ when the molecule has arrived at point 3 and starts to climb up the potential hill. It will lose kinetic energy during the climbing-up process, which equals the potential difference between points 4 and 1 in $V(x)$. After the completion of a full cycle in the cavity mode, the molecule will lose energy equal to the amount between points 1 and 2 in $V(x)$. This deceleration scheme presents an optical version of Sisyphus cooling, where the conservative motion of the molecule is interrupted by sudden transitions between high and low pump intensities. In this way, the molecule is slowed down as it travels along the standing-wave potential.

The switching of the pump in the above process can be controlled automatically by the output of the intracavity intensity via a feedback loop. The two jumps in each cycle occur at the time when the intracavity intensities are at $I_1$ and $I_2$, as shown in figure \ref{fig4}. For the case of a single molecule $I_1=0$ and $I_2=\eta_{\rm{L}}^2/\left[\kappa^2+(\Delta_c-U_0)^2\right]$ .Relevant issues of setting the two values for the deceleration of the travelling molecular packets will be discussed below.

\begin{figure}
\begin{center}
\resizebox{0.8\columnwidth}{!}{\includegraphics{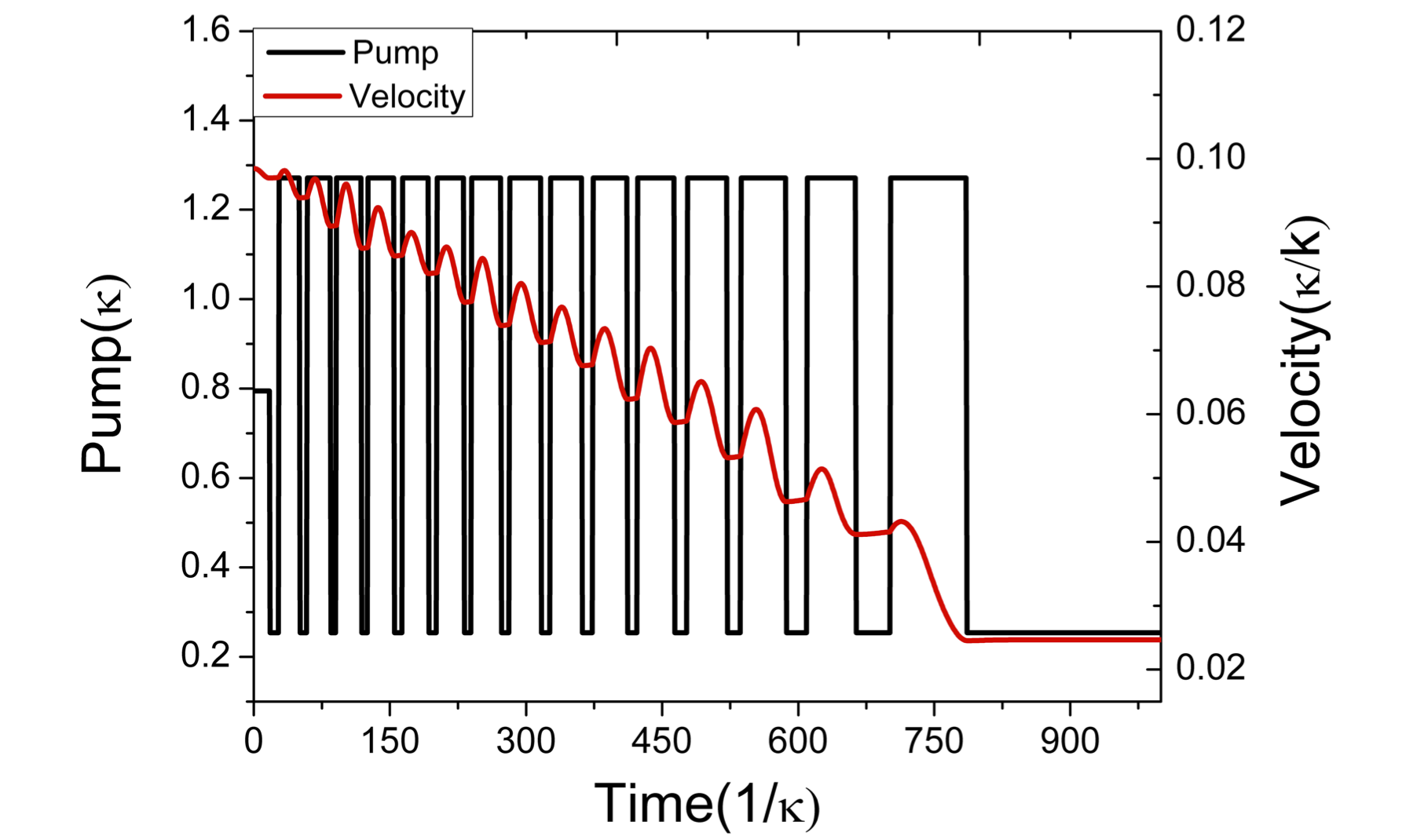}}
\end{center}
\caption{ Deceleration of the travelling molecular packets by time-varying optical pumps. The following parameters are used, $\eta_{\rm{L}}=0.8\eta_{\rm{th}}, \eta_{\rm{H}}=4\eta_{\rm{th}}, I_{\rm{min}}=100, I_{\rm{max}}=10000$, while other parameters are the same as in figure \ref{fig2}.} 
\label{fig5} 
\end{figure}

\subsection{Numerical simulations }
\label{s42}

\begin{figure}
\begin{center}
\resizebox{1.0\columnwidth}{!}{\includegraphics{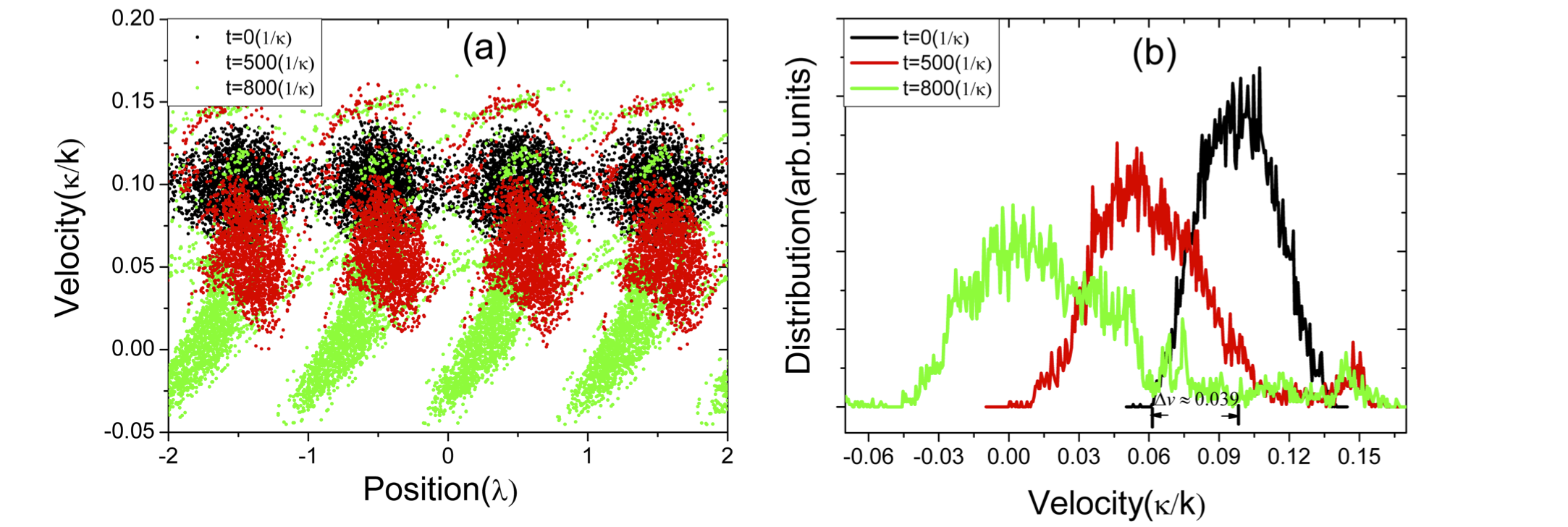}}
\end{center}
\caption{ (a) The evolution of the phase-space plots of the molecular packets at different times. (b) The velocity distributions corresponding to the phase-space plots in (a). } 
\label{fig6}
\end{figure}

When the self-consistent molecular-field steady state is achieved as discussed in section \ref{s32}, we can apply the time-varying optical pumps as described in section \ref{s41} to decelerate the bunched molecular packets. We find that in order to ensure the phase stability of the bunched molecular packets, at least one of the pump intensities should be kept above the threshold pump for phase transition as described by expression (\ref{eq11}). Figures \ref{fig5} and \ref{fig6} show the simulation results.

As seen from figure \ref{fig5}, the average velocity of the travelling molecular packets decreases linearly (constant deceleration) and the switching intervals of the pumps increase because the molecular packets spend more time in one cycle of the standing-wave cavity mode due to the reduced average velocity. The deceleration process stops at the time when the average velocity of the molecular packets is reduced to the point that a significant number of molecules no longer travel synchronously with the rest. The reduction of the molecular number of the bunched molecular packets leads to a decrease of the intracavity intensity, which will eventually be lower than the threshold $I_2$ for switching. As a result, the pump no longer switches and stays in the low intensity level as figure \ref{fig5} shows.

Figure \ref{fig6}(a) plots the position-velocity distributions of the travelling molecular packets at different times, which show the stability of the bunched molecular packets during the deceleration process. Figure \ref{fig6}(b) plots the velocity distributions corresponding to figure \ref{fig6}(a). The initial half-width of the velocity distribution of the molecular packets is $\Delta v\approx 0.039(\kappa/k)$ as marked in figure \ref{fig6}(b). The calculation based on the trapped molecules by the optical potential of the moving lattice is $\Delta v=\sqrt{2U/m}=2\eta\sqrt{\hbar \Delta_c N_{\rm{eff}}/[m(\kappa^2+\Delta_c^2)]}=0.034(\kappa/k)$, with $N_{\rm{eff}}\approx4560$ estimated from the phase-space plot ($t=0(1/\kappa)$ in figure \ref{fig6}(a)). The half-width of the velocity distribution in the deceleration process is $\Delta v\approx 0.055(\kappa/k)$, which is widened slightly (see figure \ref{fig6}(b)) compared with the initial distribution. The slight widening of the velocity distribution is accompanied by the narrowing of the position distribution due to the conservation of the phase space distribution, which is evidenced by the increased effective number of molecules $N_{\rm{eff}}\approx6700$ in the deceleration process (estimated from the simulation results). Compared to the work described in \cite{r40}, there is no extra widening factor to the velocity distribution at the end of the deceleration process in the present scheme because when the pump intensity ceases to jump between the two states, molecular velocity distribution does not spread under the adiabatic condition.

Using the single molecule approximation for the bunched molecular packets
with the effective molecular number of $N_{\rm{eff}}$, the energy extracted from the molecular
packets each cycle, as discussed in figure \ref{fig4}(b), is expressed as $ \Delta W=V(x_1)-V(x_2)\approx\hbar\Delta_c N_{\rm{eff}} (\eta_{\rm{H}}^2-\eta_{\rm{L}}^2)[\cos^2(kx_2)-\cos^2(kx_1)]/(\kappa^2+\Delta_c^2)$; since $I=\eta^2  N_{\rm{eff}}^2\cos^2(kx)/(\Delta_c^2+\kappa^2)$
and $\cos^2(kx_2)-\cos^2(kx_1)=[(\kappa^2+\Delta_c^2)/ N_{\rm{eff}} ^2] (I_{\rm{max}}/\eta_{\rm{L}}^2-I_{\rm{min}}/\eta_{\rm{H}}^2)$ the expression can be simplified to
\begin{equation}
\Delta W=\frac{\hbar \Delta_c(\eta_{\rm{H}}^2-\eta_{\rm{L}}^2)}{N_{\rm{eff}}} \left( \frac{I_{\rm{max}}}{\eta_{\rm{L}}^2}-\frac{I_{\rm{min}}}{\eta_{\rm{H}}^2} \right)
\label{eq14}
\end{equation}

Because of $I_{\rm{min}}\approx0$, $I_{\rm{max}}\propto \eta_{\rm{L}}^2$, the energy extracted in each cycle is proportional to $(\eta_{\rm{H}}^2-\eta_{\rm{L}}^2)$. This implies that the deceleration force is constant, which in turn explains a constant deceleration of the molecular packets as in figure \ref{fig5}. The deceleration is $0.85\times10^{-4}(\kappa^2/k)$ from numerical simulation (figure \ref{fig5}), while the theoretical value from equation (\ref{eq14}) is $1.1\times10^{-4}(\kappa^2/k)$, which shows good qualitative agreement considering the simple single molecule treatment.

Normally, at the end stage of the deceleration process, some molecules stop moving collectively with the molecular packets as they are decelerated to near zero velocities, so the effective number of molecules $N_{\rm{eff}}$ will decrease. Then the intracavity intensity will drop. In order to keep the jumps working at the end stage of the deceleration process, we can set the jump threshold $I_2$ lower than the maximum that can be achieved. This setup does not change the picture of the deceleration scheme but only causes the deceleration process to be not at its maximal efficiency because the energy extracted from the molecular packets each cycle is not at its maximum. Another benefit of this setup is that it will make the bunched molecular packets stay most of their time in the high pump state in each deceleration cycle, which will further guarantee the stability of the packets, because there will be not enough time for the packets to collapse during their stay in the low pump state.

\section{Composite deceleration scheme}
\label{s5}

The purpose of this section is to show that the present deceleration scheme based on time-varying laser pumps not only can work in its own right but also can be used to compensate for the reduced deceleration force at the end stage of the deceleration process as in \cite{r40}. 

When the bad cavity condition is not satisfied, the intracavity field does not respond immediately to the change in molecular distribution. This delayed response gives rise to the following two effects when the cavity changes from a bad to a good one. Firstly, it undermines the stability of the self-consistent molecular-field steady state up to a point where the collective dynamics no longer takes place. Secondly, a decelerating force on the spatially bunched molecular packets will emerge and increase correspondingly. For an effective deceleration, where phase stability and deceleration force are both needed, a compromise has to be made by choosing the appropriate cavity regime. We have recently studied such a decelerator \cite{r40} working in the intermediate cavity regime where the above two effects can be appropriately balanced. We found that the balance can be measured by the ratio $r=kv_0/\sqrt{\kappa^2+\Delta_c^2}$, where $1/kv_0$ is the time for a molecule to travel one wavelength and $1/\sqrt{\kappa^2+\Delta_c^2}$ the detuning-enhanced cavity time. A smaller (larger) $r$ means faster (slower) cavity response to the dynamics of molecules, which leads to better (poorer) spatial organization of the molecules but weaker (stronger) deceleration. By setting $r=0.3$, we demonstrated an effective deceleration.

We note that the level of nonadiabaticity depends not only on the cavity lifetime but also on the average velocity of the molecular packets. As the molecular packets are slowed down along the cavity axis as in \cite{r40} where the system approaches the adiabatic limit, the intracavity field tends to follow the changes in molecular motion better. Consequently, the deceleration force from nonadiabatic dynamics will drop. However, as discussed in section \ref{s4}, the deceleration force from time-varying optical pumps works well in the adiabatic limit, so the deceleration scheme by the time-varying optical pumps would be complementary to the deceleration scheme via nonadiabaticity of the cavity as in \cite{r40} in the sense that the latter works better than the former when the velocity of the molecular packets is higher and, vice versa, when the molecular packets have been slowed down. In the following we show that the decelerator in \cite{r40} would be more efficient when combined with the present deceleration mechanism from time-varying pumps.

\begin{figure}
\begin{center}
\resizebox{1.0\columnwidth}{!}{\includegraphics{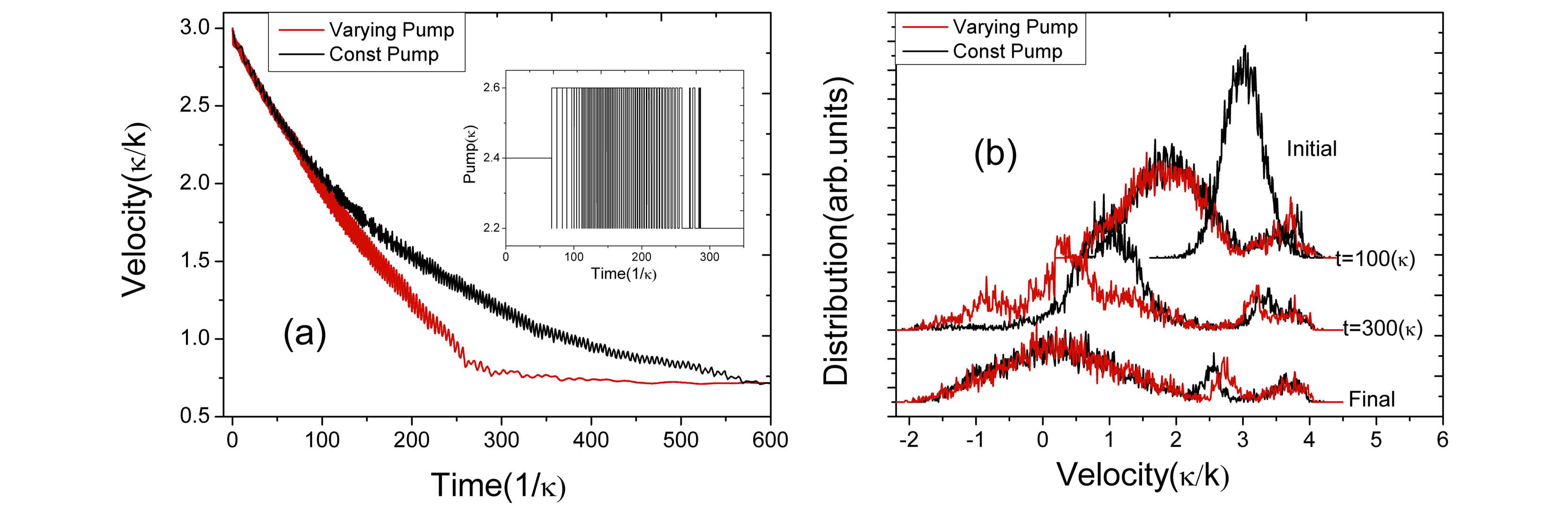}}
\end{center}
\caption{ (a) The evolution of average molecular velocity with time by constant pump (black line, which is the same as in \cite{r40}) and time-varying pumps (red). Parameters used in the simulation, $\hbar k^2/\kappa m=1.16 \times 10^{-4}, U_0=-2.28\times 10^{-5}\kappa, \Delta_c=-10\kappa, N=10^4$  and initial condition of the beam: Gaussian velocity distribution with  $k v_0/\kappa=3, k\sigma/\kappa=0.3$,   and a homogeneous spatial distribution in five wavelengths with periodic boundary condition. $\eta=2.4\kappa$  for the constant pump; $\eta_{\rm{L}}=2.2\kappa, \eta_{\rm{H}}=2.6\kappa, I_{\rm{min}}=60000, I_{\rm{max}}=120000$,  for the time-varying pumps. The insert show the pump strength with time. Panel (b) shows the velocity distributions of the molecules with the two deceleration schemes at different times in the process. } 
\label{fig7}
\end{figure}

Figure \ref{fig7}  shows the simulation results in the intermediate cavity regime (the same as in \cite{r40}) with a constant pump ($\eta=2.4\kappa$) and time-varying pumps controlled automatically by the feedback mechanism between two intensity levels ($\eta_{\rm{L}}=2.2\kappa, \eta_{\rm{H}}=2.6\kappa$). As shown in the figure, in the initial stage of the deceleration, the molecular packets move fast, so the intracavity field cannot follow the motion of the packets instantaneously. The deceleration force from time-varying laser pumps plays no role and mainly comes from the nonadiabaticity, as evidenced by the constant pump in this period shown in the inset of figure \ref{fig7}(a) and the overlapping of the two velocity curves in the initial stage. When the travelling molecular packets are slowed, the response of the intracavity field to the motion of the packets becomes better; thus the deceleration force from nonadiabaticity decreases. Meanwhile, as the system approaches the adiabatic limit, the deceleration force from time-varying pumps steps in (around $t=100(1/\kappa)$ as shown in figure \ref{fig7}(a)), which results in a constant deceleration compared with the constant pump case. Figure \ref{fig7}(b) shows the evolution of the velocity distributions of the molecules in the deceleration process. Because of constant deceleration from time-varying pumps in the adiabatic limit, the decelerator in \cite{r40} when combined with time-varying pumps is more efficient than the original proposal.

\section{Practical issues of the proposal}
\label{s6}

In this section, we will discuss the practical issues of the proposed deceleration method. To realize the proposal, the cavity-induced phase transition of a fast molecular beam and the real-time feedback control of its motion in the cavity have to be realized. The recent experimental efforts on self-organization of BEC in an optical cavity \cite{r44}, cavity-enhanced Rayleigh scatting \cite{r45} and the real-time feedback control of a single atom trajectory in a cavity \cite{r46} put our proposal within the reach of current technology.

As discussed above, a bad cavity is needed for the realization of the cavity-induced phase stability, which is different from the cavity cooling scheme where a high-finesse cavity is preferred. This is because cavity cooling needs a friction-like force from the nonadiabatic response of the cavity to condense the phase space of the cold molecular sample, whereas cavity deceleration requires an immediate response of the cavity to achieve a self-consistent molecule-field steady state. Therefore, for deceleration the cavity response time must be significantly shorter than the characteristic time of the system dynamics induced by the travelling molecular beam. For a fast molecular beam with central velocity $v_0\sim100$ m s$^{-1}$ and pump lasers with wavelength $\lambda=1\mu$m, the time for the molecular beam to travel one deceleration stage is around $t_c=\lambda/(2v_0)=5\times10^{-9}$ s, which corresponds to the rate $\kappa_c=0.2$GHz; thus a cavity with a decay rate of $\geq2$GHz would meet the bad cavity criterion ($\kappa \gg 1/t_c$) in this case. The decay rate for a Fabry-Perot cavity with length $L=1$cm and reflectance of cavity mirrors $R = 90$\% is $\kappa=c \rm{ln}(1/R)/L=3.2$GHz, which readily meets the bad cavity criterion. Self-organization of the fast molecular beam takes place around tens of nanoseconds with a cavity decay rate of $10$ GHz as can be inferred from figure \ref{fig2} in this case (note that the time needed for the self-organization also depends on the pump intensity; with the intensity at the level of the threshold pump, critical slowdown dominates). As shown in the numerical simulations earlier, the deceleration process typically requires tens of stages, so the molecular packets travel typically tens of micrometers for a duration of about 1 $\mu$s before they stop. Therefore, a cavity of length 1 cm would be sufficient theoretically for the deceleration process in our proposal. We note that in our theoretical description, transverse confinement perpendicular to the cavity axis is not discussed explicitly, but since the stopping distance (tens of micrometers) is much shorter than the transversal dimension of the mirrors (several millimetres) and, meanwhile, we use the standing-wave transverse pump scheme, where in the third remaining direction molecules are confined by the transverse envelope of the cavity and pump field as discussed in \cite{r23} we think our description is valid in this respect.

We now discuss the pump threshold power required to trigger the phase transition. For a Gaussian laser beam with waist $w_L$, the pump strength $\eta$ can be expressed in terms of the laser
power $P$ as $|\eta|=\left[ \alpha(\omega_p)/\epsilon_0\right] \sqrt{\omega_p P /(\hbar c V w_L^2\pi)}$ \cite{r26}. Substituting this expression into the pump threshold (\ref{eq11}), we have
\begin{equation}
P>m\sigma^2\frac{\delta_c^2+\kappa^2}{(-\delta_c)}\left[ \frac{\epsilon_0}{\alpha(\omega_p)}\right]^2 \left(\frac{N}{V}\right)^{-1}\frac{\pi w_L^2 c}{\omega_p}
\label{eq15}
\end{equation}
For convenience, we introduce two frequency shift parameters $r_c$ and $r_a$ to describe the pump-cavity detuning $\Delta_c=-r_c\kappa$ and the maximum shift of the empty cavity resonance frequency induced by the molecules $NU_0=-r_a\kappa$. While $r_c = r_a = 1$ is used in the cavity cooling scheme \cite{r26}, where the cavity is in resonance and thus the nonadiabatic effect is dominant, the conditions of $r_c\gg1$ and $r_a\ll1$, which essentially avoid this resonance region, are required for the effective operation of the cavity in the deceleration regime for the reasons discussed above. Normally, $r_c=5$ and $r_a=0.5$ work for the deceleration setup as we found in the numerical simulations. Using the relation $U_0=-\alpha(\omega_p)\omega_p/(\epsilon_0V)$, equation (\ref{eq15}) is simplified to $P>m\sigma^2(r_c/r_a)[\epsilon_0/\alpha(\omega_p)] w_L^2\pi c$ under the conditions of $r_c\gg1$ and $r_a\ll1$. The potential energy of a molecule in a far-off-resonant optical field in free space is $U=-2\alpha I/\epsilon_0 c$ \cite{r36}. By using the relation $I=P/w_L^2\pi$ and $m\sigma^2=k_BT/2$, the simplified pump threshold condition can be rewritten as $2\alpha I /\epsilon_0 c >k_B T (r_c/r_a)$, the meaning of which is clear compared with the case in free space: in order to trigger the cavity-induced phase transition, the potential depth generated by the intensity of the pump lasers should be larger than the transverse beam temperature modified by the cavity-related parameter $(r_c/r_a)$. We note from equation (\ref{eq15}) that the number of molecules enters only in the form of atomic density $N/V$, which shows the scaling invariance of the system as long as $N/V$ is constant. Since the coupling constant $g=\mu\sqrt{\omega_c/(2\hbar\epsilon_0 V)}$,where $\mu$ is the electric dipole transition moment, thus $N/V\propto Ng^2$, which implies that the smaller coupling constant with a bigger cavity can be compensated by increasing the molecular number.

The validity of the pump threshold (\ref{eq15}) can be tested by the experimental data from \cite{r44} where the self-organization-like phase transition has been demonstrated with $^{87}$Rb BEC. The parameters used in the experiment are $(g, \kappa, \gamma)=2\pi\times(10.6, 1.3, 3.0)$MHz, cavity length of 178 $\mu$m, waist radius of 25 $\mu$m, pump-atom detuning $\Delta_a$ of 4.3 nm from the atomic $D_2$ line ($\Delta_a\approx 2100$GHz) and $NU_0=-6.5\kappa$, where $U_0=g^2/\Delta_a$\cite{r23, r24}in this case. If we choose one set of parameters $(\Delta_c, P)=(-2\pi\times20$MHz, $400\mu$W) from their phase diagram and using the BEC temperature of $\sim100$nK and the pump area of $70\mu$m $\times70\mu$m, the calculated pump threshold power according to expression (\ref{eq15}) is $\sim250\mu$W, which is close to the real experimental value of $400\mu$W. A further simple extrapolation of the above results to a fast $^{87}$Rb gas beam with a transverse temperature of $100$ mK in a bad cavity ($r_c/r_a=10$) would require a pump threshold power of $2$ kW. Currently, single mode ytterbium-doped fibre lasers with an output power of $2$ kW are commercially available and some $5$ kW has been demonstrated in the laboratory environment \cite{r47}.

We then consider an example of benzene molecules, which have been used in the experiment on single-stage optical Stark deceleration \cite{r36}. The molecule has an average polarizability of $\alpha(\omega_p)=11.6\times10^{-40}$ Cm$^2  $$V^{-1}$ at the pump laser wavelength of $\lambda=1064$nm. For a pulsed fast beam of molecular benzene with velocity spread of $\sigma=10$ m s$^{-1}$, pump laser waist (also the molecular beam length) of 1 mm and the cavity-related parameter $r_c/r_a=10$, the required pump threshold intensity would be $I >9.3\times 10^{10}$  W cm$^{-2}$, which can be readily achieved by the pulsed laser used in \cite{r36} if stretching it to a microsecond-duration one. Since the pump detuning is of the order of $10^{14}$ Hz, the saturation parameter (to be discussed in the following) is calculated to be 0.01\%, which is far below the acceptable level of 1\% where the population excitation and thus spontaneous emission are negligible. The use of a pulsed laser is justified by the short time scale, typically within 1 $\mu$s, of the deceleration process. We have discussed the use of a single laser pulse with microsecond duration for the pump in our previous work \cite{r40}. To produce a time-varying optical pump field in the present deceleration scheme, the pulsed laser can be modulated by electro-optic modulators \cite{r48} or fibre modulators \cite{r25} with bandwidths at tens of GHz, which are much larger than the required cavity decay rate of several GHz. The molecular density in this case is about $N/V\approx10^{15}$cm$^{-3}$, and such an intense low- energy molecular beam can be readily obtained by the method of Ôpressure shockÕ as described in \cite{r49}.

Since the pump lasers are far-off-resonance from all electronic transitions in the above analysis, we can safely neglect spontaneous emissions in the analysis \cite{r44, r45}. However, the operation conditions can be significantly relaxed if one makes use of the effects of resonantly enhanced dipole moment. The parameter setting in this case depends on the chosen molecule and the number of open transitions; however, if the pump frequency detuning from the transitions is much larger than the energy splitting of the allowed transitions, then an approximated two-level model is valid \cite{r28, r40}. In order to suppress spontaneous emission, the saturation parameter $s=|\alpha|^2g^2/\Delta_a^2$ \cite{r24} should be negligible in this case, where $|\alpha|^2$ is the intracavity photon number. By inserting the expression (\ref{eq12}) for $|\alpha|^2$ and after some algebra, the saturation parameter is simplified to $s\approx m\sigma^2(r_a/r_c)\hbar\Delta_a$, which means that in order to avoid significant population excitation, the energy associated with the detuning should be much larger than the transverse temperature of the beam modified by the cavity-related parameter $(r_c/r_a)$. Such an operation has been discussed in detail in our previous work \cite{r40}, which proves to be feasible by using a pump source far-detuned from the allowed optical transitions (hundreds to thousands of GHz) to suppress spontaneous emission. Recently, a ``supersonic electric conveyor belt" experiment, where metastable CO molecules are trapped and transported in travelling potential wells at constant velocities on a chip, has been demonstrated \cite{r50}. Our self-consistent molecule-field steady state described in section \ref{s3} can be taken as a cavity-based version of this transportation experiment. We consider the pump threshold power to realize this ``conveyor belt" experiment with a bad cavity of $r_c/r_a = 10$. For a pulsed beam (1 mm long) of CO molecules at a transverse temperature of 20 mK with $Q_2(1)$ transition of $a^3\Pi\leftarrow X^1\Sigma^+$, which has a transition dipole moment of 1.37 D, as used in \cite{r50}, to avoid significant population excitation ($s\sim0.01$), the detuning is then $\Delta_a\sim1.3\times 10^{10}$Hz. The pump threshold power in this case is about 1 kW, which is within reach for experiments \cite{r47}. Other methods to reduce the pump threshold power, such as seeding the cavity and pump power recycling with a second cavity, are also available \cite{r20, r26}.

\section{Conclusions }
\label{s7}

In this paper, we have proposed a new scheme for decelerating a fast molecular beam in the bad cavity regime based on the cavity-induced phase stability mechanism and time-varying optical pumps. We first explored in detail the dynamical interplay between a fast molecular beam moving along the axis of a standing-wave cavity and the intracavity field formed from the scattering of the transversal pump photons by the molecules in the beam. We found that in the adiabatic limit, a phase transition, from which travelling molecular packets are formed from the initial spatially homogeneous fast molecular beam above some threshold pump, results in a well-defined self-consistent molecule-field steady state that can be adopted for multistage deceleration. This phase stability mechanism from the cavity-induced collective behaviour of molecules ensures the phase stability of the majority of the molecules in the cavity rather than a fraction (small acceptance volume) as in phase space filtering techniques. We subsequently introduced the deceleration force to extract energy from the molecular packets in a similar way to electrostatic Stark deceleration by introducing sudden switching between two levels of the pump intensities, which are synchronous with the up and down processes of the molecular packets in the standing-wave potential. However, in our scheme the switching sequence of the pumps is achieved automatically by feedback, rather than timed externally as in electrostatic Stark deceleration. Due to no extra widening factor to the velocity distribution at the end of the deceleration process compared with \cite{r40}, the present proposal can maintain the low transverse temperature and high density of the beam while it requires only tens of deceleration stages. Practical issues in realizing the proposal are also discussed in detail.

An important difference between our method and electrostatic Stark deceleration is that while the phase stability and deceleration in the latter are interwoven, i.e. at higher deceleration rates, only smaller numbers of stably bunched molecules can be handled and vice versa \cite{r33}, our method allows the engineering of phase stability and deceleration force separately. Our method is also different from both the single and multi-stage optical Stark decelerators in free space based on optical lattice as demonstrated or proposed previously \cite{r36, r38}. For the experimentally demonstrated single-stage optical Stark decelerator for molecules with pulsed optical lattice \cite{r36}, since there is no bunching effect due to the single-time interaction of the molecules with the laser, the width of the slowed molecules in the velocity space is quite broad. For the multistage optical Stark decelerator in free space as proposed in \cite{r48}, where the bunching effect is present in obtaining a narrowed velocity distribution, the energy extracted from the molecular packet for each period is small due to the lack of cavity-induced collective effect, and the number of stages required for the deceleration process is tens of thousands of stages \cite{r48}. By using the collective enhancement effect with a cavity, the deceleration process needs only tens of stages while keeping the initial low transverse temperature and high density of the beam. We note that the work described in \cite{r25} was based on the assumption that a molecular sample below 1 K has been prepared by using a decelerator technique, before both external and internal degrees of freedom of these molecules can be further cooled by a cavity. Our proposed method can serve this purpose by providing a high-density molecular beam at the required temperature. Since deceleration or cooling of external motion is a relatively fast stage, in which the internal motion is not affected by the scattering process, the present work together with previous cavity cooling studies \cite{r20, r21, r22, r23, r24, r25, r26, r27, r28} shows the feasibility of bringing a hot molecular beam into the ultracold regime with only a cavity setup where high density and low temperature can be achieved at the same time in principle. We hope this study will stimulate experimental efforts toward this exciting new possibility. Finally we would like to point out that while this paper has focused only on a deceleration scheme based on the self-consistent molecule-field steady state, this steady state may be used for other applications as well, such as the supersonic conveyor belt \cite{r50}.

\section*{Acknowledgements}

ZL acknowledges financial support from Scottish Universities Physics Alliance (SUPA).


\vspace{1cm}

\begin{thebibliography}{99}

\bibitem{r1} Anderson M H, Ensher J R, Matthews M R, Wieman C E and Cornell E A 1995 {\it ObservationofBose-Einstein condensation in a dilute atomic vapour Science} {\bf 269} 198Ð201
\bibitem{r2}  O'Hara K M, Hemmer S L, Gehm M E, Granade S R and Thomas J E 2002 {\it Observation of a strongly interacting degenerate Fermi gas of atoms} Science {\bf 298} 2179-2182
\bibitem{r3} Giorgini S, Pitaevskii L P and Stringari S 2008 {\it Theory of ultracold atomic Fermi gases} Rev. Mod. Phys. {\bf 80} 1215-1274
\bibitem{r4} Greiner M, Mandel O, Esslinger T, Hansch T W and Bloch I 2002 {\it Quantum phase transition from a superfluid to a Mott insulator in a gas of ultracold atoms} Nature {\bf 415} 39-44
\bibitem{r5} Carr L D, DeMille D, Krems R V and Ye J 2009 {\it Cold and ultracold molecules: science, technology and applications}  New J. Phys. {\bf 11} 055049
\bibitem{r6} Baranov M A 2008 {\it Theoretical progress in many-body physics with ultracold dipolar gases} Phys. Rep. {\bf 464} 71-111
\bibitem{r7} DeMille D 2002 {\it Quantum computation with trapped polar molecules} Phys. Rev. Lett. {\bf 88} 067901
\bibitem{r8} Hudson E R, Lewandowski H J, Sawyer B C and Ye J 2006 {\it Cold molecule spectroscopy for constraining the evolution of the fine structure constant} Phys. Rev. Lett. {\bf 96} 143004
\bibitem{r9} Quack M, Stohner J and Willeke M 2008 {\it High-resolution spectroscopic studies and theory of parity violation in chiral molecules} Annu. Rev. Phys. Chem. {\bf 59} 741-769
\bibitem{r10} Hudson J J, Sauer B E, Tarbutt M R and Hinds E A 2002 {\it Measurement of the electron electric dipole moment using YbF molecules} Phys. Rev. Lett. {\bf 89} 023003
\bibitem{r11} van de Meerakker S Y T, Vanhaecke N, van der Loo M P J, Groenenboom G C and Meijer G 2005 {\it Direct measurement of the radiative lifetime of vibrationally excited OH radicals} Phys. Rev. Lett. {\bf 95} 013003
\bibitem{r12} Campbell W C, Groenenboom G C, Lu H I, Tsikata E and Doyle J M 2008 {\it Time-domain measurement of spontaneous vibrational decay of magnetically trapped NH} Phys. Rev. Lett. {\bf 100} 083003
\bibitem{r13} Krems R V 2008 {\it Cold controlled chemistry} Phys. Chem. Chem. Phys. {\bf 10} 4079-4092
\bibitem{r14} Shuman E S, Barry J F, Glenn D R and DeMille D 2009 {\it Radiative force from optical cycling on a diatomic molecule} Phys. Rev. Lett. {\bf 103} 223001; Shuman E S, Barry J F and DeMille D 2010 {\it Laser cooling of a diatomic molecule} Nature {\bf 467} 820-823
\bibitem{r15} Jones K M, Tiesinga E, Lett P D and Julienne P S 2006 {\it Ultracold photoassociation spectroscopy: Long-range molecules and atomic scattering} Rev. Mod. Phys. {\bf 78} 483-535
\bibitem{r16} Kohler T, Goral K and Julienne P S 2006 {\it Production of cold molecules via magnetically tunable Feshbach resonances} Rev. Mod. Phys. {\bf 78} 1311-1361
\bibitem{r17} Ni K K, Ospelkaus S, de Miranda M H G, Pe'er A, Neyenhuis B, Zirbel J J, Kotochigova S, Julienne P S, Jin D S and Ye J 2008 {\it A high phase-space-density gas of polar molecules} Science {\bf 322} 231-235
\bibitem{r18} Weinstein J D, de Carvalho R, Guillet T, Friedrich B and Doyle J M 1998 {\it Magnetic trapping of calcium monohydride molecules at millikelvin temperatures} Nature {\bf 395} 148-150
\bibitem{r19} Doret S C, Connolly C B, Ketterle W and Doyle J M 2009 {\it Buffer-Gas Cooled Bose-Einstein Condensate} Phys. Rev. Lett. {\bf 103} 103005
\bibitem{r20} Lev B L, Vukics A, Hudson E R, Sawyer B C, Domokos P, Ritsch H and Ye J 2008 {\it Prospects for the cavity-assisted laser cooling of molecules} Phys. Rev. A {\bf 77} 023402
\bibitem{r21} Horak P, Hechenblaikner G, Gheri K M, Stecher H and Ritsch H 1997 {\it Cavity-induced atom cooling in the strong coupling regime} Phys. Rev. Lett. {\bf 79} 4974-4977
\bibitem{r22} Vuletic V and Chu S 2000 {\it Laser cooling of atoms, ions, or molecules by coherent scattering} Phys. Rev. Lett. {\bf 84} 3787-3790    
\bibitem{r23} Domokos P and Ritsch H 2002 {\it Collective cooling and self-organization of atoms in a cavity} Phys. Rev. Lett. {\bf 89} 253003
\bibitem{r24} Horak P and Ritsch H 2001 {\it Scaling properties of cavity-enhanced atom cooling} Phys. Rev. A {\bf 64} 033422; Asboth J K, Domokos P, Ritsch H and Vukics A 2005 {\it Self-organization of atoms in a cavity field: Threshold, bistability, and scaling laws} Phys. Rev. A {\bf 72} 053417 
\bibitem{r25} Morigi G, Pinkse P W H, Kowalewski M and de Vivie-Riedle R 2007 {\it Cavity cooling of internal molecular motion} Phys. Rev. Lett. {\bf 99} 073001; 2007 {\it Cavity cooling of translational and ro-vibrational motion of molecules: ab initio-based simulations for OH and NO} Appl. Phys. B {\bf 89} 459-467
\bibitem{r26} Salzburger T and Ritsch H 2009 {it Collective transverse cavity cooling of a dense molecular beam} New J. Phys. {\bf 11} 055025 
\bibitem{r27} Lu W, Zhao Y and Barker P F 2007 {\it Cooling molecules in optical cavities} Phys. Rev. A {\bf 76} 013417
\bibitem{r28} Zhao Y K, Lu W P, Barker P F and Dong G J 2009 {\it Self-organisation and cooling of a large ensemble of particles in optical cavities} Faraday Discuss. {\bf 142} 311-318
\bibitem{r29} Chan H W, Black A T and Vuletic V 2003 {\it Observation of collective-emission-induced cooling of atoms in an optical cavity} Phys. Rev. Lett. {\bf 90} 063003; 2003 {\it Observation of collective friction forces due to spatial self-organization of atoms: from Rayleigh to Bragg scattering} Phys. Rev. Lett. {\bf 91} 203001
\bibitem{r30} Maunz P, Puppe T, Schuster I, Syassen N, Pinkse P W H and Rempe G 2004 {\it Cavity cooling of a single atom} Nature {\bf 428} 50-52
\bibitem{r31} Nagorny B, Elsasser T and Hemmerich A 2003 {\it Collective atomic motion in an optical lattice formed inside a high finesse cavity} Phys. Rev. Lett. {\bf 91} 153003
\bibitem{r32} Bethlem H L, Berden G and Meijer G 1999 {\it Decelerating neutral dipolar molecules} Phys. Rev. Lett. {\bf 83} 1558-1561; Bethlem H L, Berden G, Crompvoets F M H, Jongma R T, van Roij A J A and Meijer G 2000 {\it Electrostatic trapping of ammonia molecules} Nature {\bf 406} 491-494
\bibitem{r33} Gubbels K, Meijer G and Friedrich B 2006 {\it Analytic wave model of Stark deceleration dynamics} Phys. Rev. A {\bf 73} 063406
\bibitem{r34} Narevicius E, Libson A, Parthey C G, Chavez I, Narevicius J, Even U and Raizen M G 2008 {\it Stopping supersonic beams with a series of pulsed electromagnetic coils: An atomic coilgun} Phys. Rev. Lett. {\bf 100} 093003; 2008 {\it Stopping supersonic oxygen with a series of pulsed electromagnetic coils: A molecular coilgun} Phys. Rev. A {\bf 77} 051401
\bibitem{r35} Hogan S D, Wiederkehr A W, Schmutz H and Merkt F 2008 {\it Magnetic trapping of hydrogen after multistage Zeeman deceleration} Phys. Rev. Lett. {\bf 101} 143001
\bibitem{r36} Fulton R, Bishop A I and Barker P F 2004 {\it Optical Stark decelerator for molecules} Phys. Rev. Lett. {\bf 93} 243004; Fulton R, Bishop A I, Shneider M N and Barker P F 2006 {\it Controlling the motion of cold molecules with deep periodic optical potentials} Nat. Phys. {\bf 2} 465-468 
\bibitem{r37} Averbukh I S and Prior Y 2005 {\it Laser cooling in an optical shaker} Phys. Rev. Lett. {\bf 94} 153002
\bibitem{r38} Enomoto K and Momose T 2005 {\it Microwave Stark decelerator for polar molecules} Phys. Rev. A {\bf 72} 061403(R); Kuma S and Momose T 2009 {\it Deceleration of molecules by dipole force potential: a numerical simulation} New J. Phys. {\bf 11} 055023
\bibitem{r39} Vilensky M Y, Prior Y and Averbukh I S 2007 {\it Cooling in a bistable optical cavity} Phys. Rev. Lett. {\bf 99} 103002
\bibitem{r40} Lan Z H, Zhao Y K, Barker P F and Lu W P 2010 {\it Deceleration of molecules in a supersonic beam by the optical field in a low-finesse cavity} Phys. Rev. A {\bf 81} 013419
\bibitem{r41} Griesser T, Ritsch H, Hemmerling M and Robb GRM 2010 {\it A Vlasov approach to bunching and selfordering
of particles in optical resonators} Eur. Phys. J. D {\bf 58} 349Ð68
\bibitem{r42} Nagy D,Szirmai G and Domokos P 2008 {\it Self-organization of a Bose-Einstein condensate in an optical cavity}
Eur. Phys. J. D {\bf 48} 127-37
\bibitem{r43} Maes C, Asboth J K and Ritsch H 2007 {\it Self ordering threshold and superradiant backscattering to slow a fast gas beam in a ring cavity with counter propagating pump} Opt. Express {\bf 15} 6019-6035
\bibitem{r44} Baumann K, Guerlin C, Brennecke F and  Esslinger T 2010 {\it Dicke quantum phase transition with a superfluid gas in an optical cavity} Nature {\bf 464} 1301-U1 
\bibitem{r45} Motsch M, Zeppenfeld M, Pinkse P W H and Rempe G 2010 {\it Cavity-enhanced Rayleigh scattering} New J. Phys. {\bf 12} 063022 
\bibitem{r46} Kubanek A, Koch M, Sames C, Ourjoumtsev A, Pinkse P W H, Murr K and Rempe G 2009 {\it Photon-by-photon feedback control of a single-atom trajectory} Nature {\bf 462} 898-901
\bibitem{r47} Maran J N, Jeong Y, Yoo S, Sahu J and Nilsson J 2008 {\it Progress in high-power single frequency master oscillator power amplifier} Photonics North {\bf 7099} X990
\bibitem{r48} Yin Y L, Zhou Q, Deng L Z, Xia Y and Yin J P 2009 {\it Multistage optical Stark decelerator for a pulsed supersonic beam with a quasi-cw optical lattice} Opt. Express {\bf 17} 10706-10717
\bibitem{r49} Makarov G N 2002 {\it Generation of intense low-energy molecular beams} Chem. Phys. Lett. {\bf 366} 490-495
\bibitem{r50} Meek S A, Bethlem H L, Conrad H and Meijer G 2008 {\it Trapping molecules on a chip in traveling potential wells} Phys. Rev. Lett. {\bf 100} 153003

\end{thebibliography}
\end{document}